\documentclass[preprint]{iucr}              
\usepackage{graphicx}
\usepackage{xspace}
\usepackage{xcolor}
\usepackage{amsmath, amsthm, amssymb, amsfonts}

     \journalcode{B}              

\usepackage{amsmath}
\usepackage{bm}
\usepackage{longtable}
\usepackage{braket}

\newcommand{\mm}[1]     {\ifmmode {#1} \else{}${#1}$\fi}
\newcommand{\mmm}[1]    {\ifmmode{}#1 \else{}${#1}$\fi}
\newcommand{\beq}[1]{\begin{equation}\label{#1}}
\newcommand{\eeq}{\end{equation}}

\def \mno{{\rm MnO}}
\def \nio{{\rm NiO}}

\def\vec#1{\mm{{\rm\bm{{\mathrm#1}}}}}
\def\kI{\mm{\vec{k}_1}}
\def\kII{\mm{[{1\over2} {1\over2} {1\over2}]}}
\def\kIIp{\mm{({1\over2} {1\over2} {1\over2})}}
\def\kIIR{\mm{[{0} {0} {3\over2}]}}

\newcommand{\minus}{\scalebox{0.5}[1.0]{$-$}}
\def\kstar{\mm{\{[{1\over2} {1\over2} {1\over2}],}
                        \mm{[\minus{1\over2} {1\over2} {1\over2}],}
                        \mm{[{1\over2} \minus{1\over2} {1\over2}],}
                        \mm{[{1\over2} {1\over2} \minus{1\over2}]\}}}
\def\kstarp{\mm{\{[{1\over2} {1\over2} {1\over2}],}
                        \mm{[0 0 {1\over2}],}
                        \mm{[0 {1\over2} 0],}
                        \mm{[{1\over2} 0 0 ]\}}}                        

\def\Fm{\mm{Fm\bar{3}m}}
\def\Rm{\mm{R\bar{3}m}}
\def\RIc{\mm{R_I\bar{3}c}}

\def\Cc{\mm{C_c2/c}}

\def\P6m{\mm{P{6}/m}}

\def \ot{{\mm{{1\over2}}}}

 \def\figsiz{8cm}
\def\figsizcm{8}

\begin{document}                  



\title{On Magnetic and Crystal Structures of {\bm\nio} and {\bm\mno}} 
\shorttitle{Magnetic and Crystal Structures of {\bm\nio} and {\bm\mno}}
\author{V.}{Pomjakushin}  

\aff{Laboratory for Neutron Scattering and Imaging (LNS), Paul Scherrer
Institut (PSI), CH-5232 Villigen PSI \country{Switzerland}}









\maketitle                        
\begin{synopsis}
The magnetic and crystal structures of manganese and nickel monoxides have been analysed for possible multi-$k$ solutions and compared with high-resolution neutron diffraction data.
\end{synopsis}

\begin{abstract}

The magnetic and crystal structures of manganese and nickel monoxides have been studied by high-resolution neutron diffraction.  
The known 1$k$-structures based on the single propagation vector \kII\ for the parent paramagnetic space group \Fm\ are forced to have monoclinic magnetic symmetry and are not possible in rhombohedral symmetry. 
However, the monoclinic distortions from the rhombohedral \Rm-metric allowed by symmetry are very small, and the explicit monoclinic splittings of the diffraction peaks have not been experimentally observed.  
We analyze the magnetic crystallographic models metrically compatible with our experimental data in full detail by using the isotropy subgroup representation approach, including rhombohedral solutions based on the propagation vector star \kstar. Although the full star rhombohedral \RIc-structure can equally well fit our diffraction data for \nio, we conclude that the best solution for the crystal and magnetic structures for \nio\ and MnO is the 1$k$-monoclinic model with the magnetic space group $C_c2/c$ (BNS 15.90, UNI symbol $C2/c.1'_c[C2/m]$).

\end{abstract}

\section{Introduction}
Since the first neutron diffraction studies of manganese and nickel monoxides \cite{RN547}, interest in the magnetic and crystal structures of \mno\ and \nio\ has been maintained, 
see e.g. experimental studies \cite{RN527,RN571}  and theoretical first-principles calculations \cite{RN499,RN572} and references cited therein. 
It is well-established experimentally that the cubic paramagnetic crystal structure \Fm\ transforms metrically to a rhombohedral one \Rm\ at the transition to an antiferromagnetically (AFM) ordered structure below the N\'eel temperature $T_N$. The AFM structure is  well described  with one propagation vector $\kI=\kII$ (in cubic metric) and the moment directed perpendicular to $\vec{k}_1$.  
This type of structure corresponds to the monoclinic magnetic space group (MSG) $C_c2/m$ (12.63 BNS), whose properties and relation to the parent cubic group are shown in Table~\ref{tab_str}. 
The second high-symmetry monoclinic MSG is $C_c2/c$ (15.90) and it allows spin components both parallel and perpendicular to \kI\ (the $x$-component in $C_c2/c$-settings is perpendicular to cubic \kI, and $z$-component makes an angle of about 35~degrees  to \kI-direction). Experimentally, one does see rhombohedral distortion in the titled compounds, but the monoclinic distortions from the \Rm\ metric allowed by symmetry are very small and the explicit monoclinic splittings of the diffraction peaks have not been experimentally observed. Thus, a quite logical idea would be to try to find solutions in rhombohedral symmetry that would fit the experimental data. The only rhombohedral MSG allowing non-zero magnetic moments on Ni/Mn for a one-\kI\ structure is \RIc(1k) with a single Wyckoff position for Mn, but this structure cannot fit experimental diffraction intensities at all because it gives zero intensity for the most intense magnetic peak $(\ot,\ot,\ot)$. However, the rhombohedral structures based on several propagation vectors (arms) from the propagation vector star $\{\vec{k}_1\}=$ \kstar\ could, in principle, be compatible with experimental data. In this paper we study possible rhombohedral and monoclinic solutions for multi-$k$ magnetic structures based on several arms from the star $\{\kI\}$ and compare the models with high-resolution neutron diffraction data.  

\section{Experimental details}
Neutron powder diffraction measurements were carried out at the high-resolution HRPT diffractometer \cite{hrpt} at the SINQ neutron spallation source (PSI, Switzerland), using high-resolution mode ($\delta d/d>10^{-3}$) for neutron wavelength $\lambda=1.155$~\AA\ and high-intensity mode for  $\lambda=1.886$~\AA.   The refinements of the structure parameters were done using the {\tt FULLPROF}\cite{Fullprof} program, with the use of its internal tables for neutron scattering lengths. The symmetry analysis was performed using the {\tt ISODISTORT} tool based on the {\tt ISOTROPY} software \cite{isot,isod}, the software tools of the Bilbao crystallographic server \cite{Bilbao}, and the {\tt BasiRep} program \cite{Fullprof}.  The magnetic structure figures were prepared using the {\tt VESTA} \cite{VESTA} software program.

\section{Symmetry analysis}
\begin{figure}
  \begin{center}
    \includegraphics[width=\figsiz]{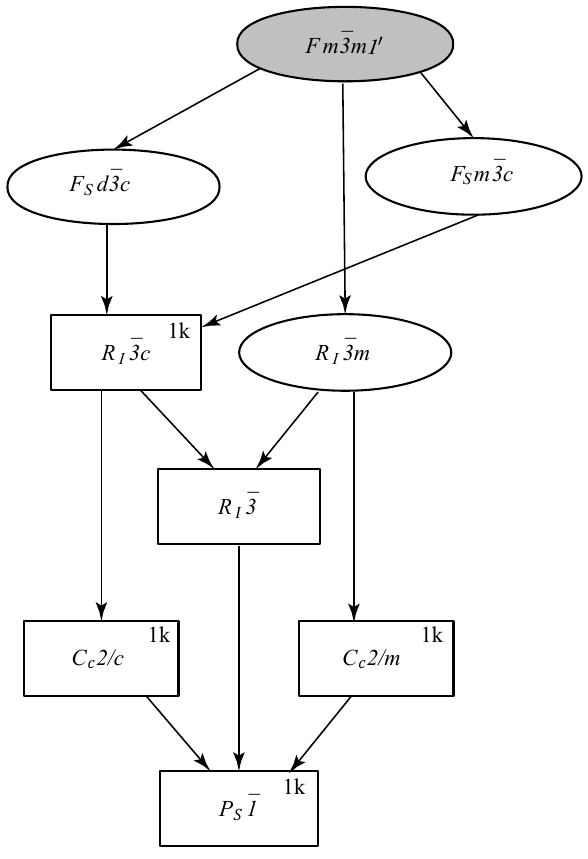}
  \end{center}
\caption
{Group-subgroup graph for 4$k$ structures based on the propagation vector star generated by $[{1\over2} {1\over2} {1\over2}]$ in the \Fm1' space group with the magnetic atom in 4a (0,0,0) position.
The possible 1$k$ structures are indicated by label 1$k$. Only cubic, rhombohedral and monoclinic groups are shown, whose crystal metrics are compatible with the experiment.}
 \label{k_graph}
\end{figure}

The paramagnetic crystal structure of \mno\ and \nio\  is $Fm\bar{3}m$  with Mn or Ni in the 4a (0,0,0) position and oxygen in  the 4b (1/2,1/2,1/2) position.  All magnetic peaks are indexed with the propagation vector \kI=\kII, which is the L-point of the Brillouin zone (BZ).  The decomposition of the magnetic representation for the 4a position into irreducible representations (irreps) reads $mL_2^{+}(\tau_3) \oplus mL_3^{+}(\tau_5)$ with the dimensions of irreps being 1D and 2D, respectively. The nomenclature for the irreps is given according to \cite{isod,Bilbao} with Kovalev's notation in parentheses.  For the single propagation vector 1$k$ structure, there are four subgroups shown in Fig.~\ref{k_graph}. The one-dimensional irrep $mL_2^+$ results in the magnetic structure $R_I\bar{3}c$ with the moments parallel to \kI. The subgroups generated by the irrep $mL_3^+$ as the primary order parameter result in the magnetic structure with the moments perpendicular to \kI\ in MSG $C_c2/m$, whereas two other groups $C_c2/c$ and $P_S\bar{1}$ allow the presence of irrep $mL_2^+$ as a secondary order parameter, which allows a spin component along the propagation vector.

We note that for the 1$k$ case,  $C_c2/m$ and $R_I\bar{3}c$ are maximal magnetic subgroups of the parent grey Shubnikov group $\Fm{1'}$. 
The basis transformation from cubic to monoclinic cells for both  $C_c2/m$ and $C_c2/c$ reads: (1/2,1/2,-1), (1/2,-1/2,0), (-1,-1,0). In MSG $C_c2/c$, Ni and O are in the 4c $(000 | m_x,0,m_z)$ and 4b $(0{1\over2}{1\over4})$ positions. In MSG $C_c2/m$, Ni and O are in the 4a $(000 | 0,m_y,0)$ and 4d $(0{1\over2}{1\over4})$ positions, respectively. The $(ab)$ monoclinic plane is perpendicular to the propagation vector \kI, so in $C_c2/m$ the allowed spin direction is only perpendicular to \kI. The propagation vector \kI, $a$, and $c$ axes lie in the same plane with an ideal pseudocubic monoclinic angle (between $a$ and $c$) $\beta=\arccos(-1/\sqrt{3})$ $\simeq 125.264^\circ$, and an angle between \kI\ and the $c$-axis $270-\beta\simeq 144.736^\circ$.

For the multi-$k$ structures of the propagation vector star \kstar\ of \Fm, the number of possible subgroups is significantly larger. In the primitive basis, the arms of the star are \kstarp. As we will show below from the diffraction experiments, the crystal metric in the AFM state is rhombohedral. For that reason we exclude possible tetragonal and orthorhombic MSG and their subgroups because the crystal metric from the experiment cannot be fitted to the orthorhombic one.  The eight MSG that are left to consider are shown as a tree graph in Fig.~\ref{k_graph}.  Four subgroups labeled with 1$k$ are MSG allowed for a single propagation vector \kI. We note that although the labels for the groups for 1$k$-, 2$k$-, 3$k$-, and 4$k$-cases are the same, the magnetic structures are different. Three rhombohedral MSG are $R_I\bar{3}m$ ($mL_3^{+}$, P22),  $R_I\bar{3}c$ ($mL_3^{+}$, P21), and $R_I\bar{3}$ ($mL_3^{+}$, C56), with the active primary irrep and direction of the order parameter shown in parentheses according to \cite{isod}.
The basis transformation is (-1,0,-1), (0,1,1), (2,2,-2), 
and the single 4a site in \Fm\ is split into 18e and  6a positions with spin components 18e (mx,0,0), 6a (0,0,0); 18e (mx,2mx,mz), 6a (0,0,mz); and 18e(mx,my,mz), 6a (0,0,mz) for the above-listed rhombohedral groups, respectively. The two last MSG allow the presence of the secondary irrep $mL_2^{+}$. 
The first MSG does not allow a non-zero magnetic moment on all magnetic atom sites and hence can be excluded.
Interestingly, the MSG $R_I\bar{3}c$ generated by the primary irrep $mL_3^{+}$ with three active k-vectors can be also generated by the irrep $mL_2^{+}$, C7 with four k-vectors. The magnetic structures in both cases are the same. The MSG $R_I\bar{3}$ is a direct subgroup of $R_I\bar{3}c$ with identity basis transformation and has the same positions for Mn/Ni and O atoms, but due to the loss of the mirror plane the point group allows a general direction of the magnetic moment in the 18e position. 

The full star monoclinic subgroup $C_c2/c$ ($mL_2^{+}$, S2) is constructed in a similar way to the 1$k$-$C_c2/c$, but the single 4c position is split into three special positions:  4c (mx,0,mz), 4d (mx,0,mz), and 8f (mx,my,mz). If the spin components are constrained to be the same for all three positions, then the magnetic 4$k$-structure is equivalent to the 1$k$ structure.

\section{Neutron Diffraction Results}

\subsection{Crystal structure}
\label{crys_sec}
\begin{figure}
  \begin{center}
    \includegraphics[width=\figsiz]{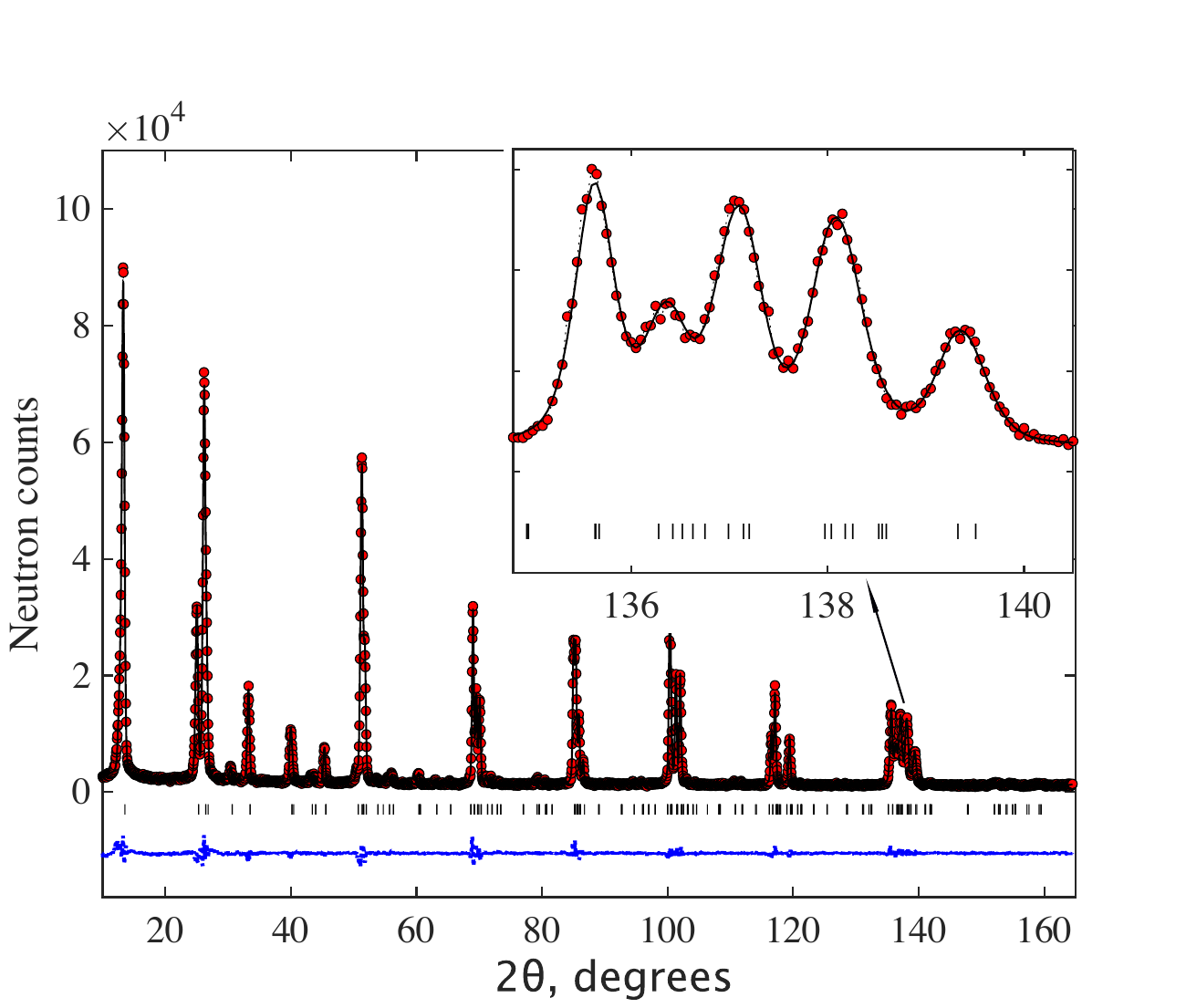}
    \includegraphics[width=\figsiz]{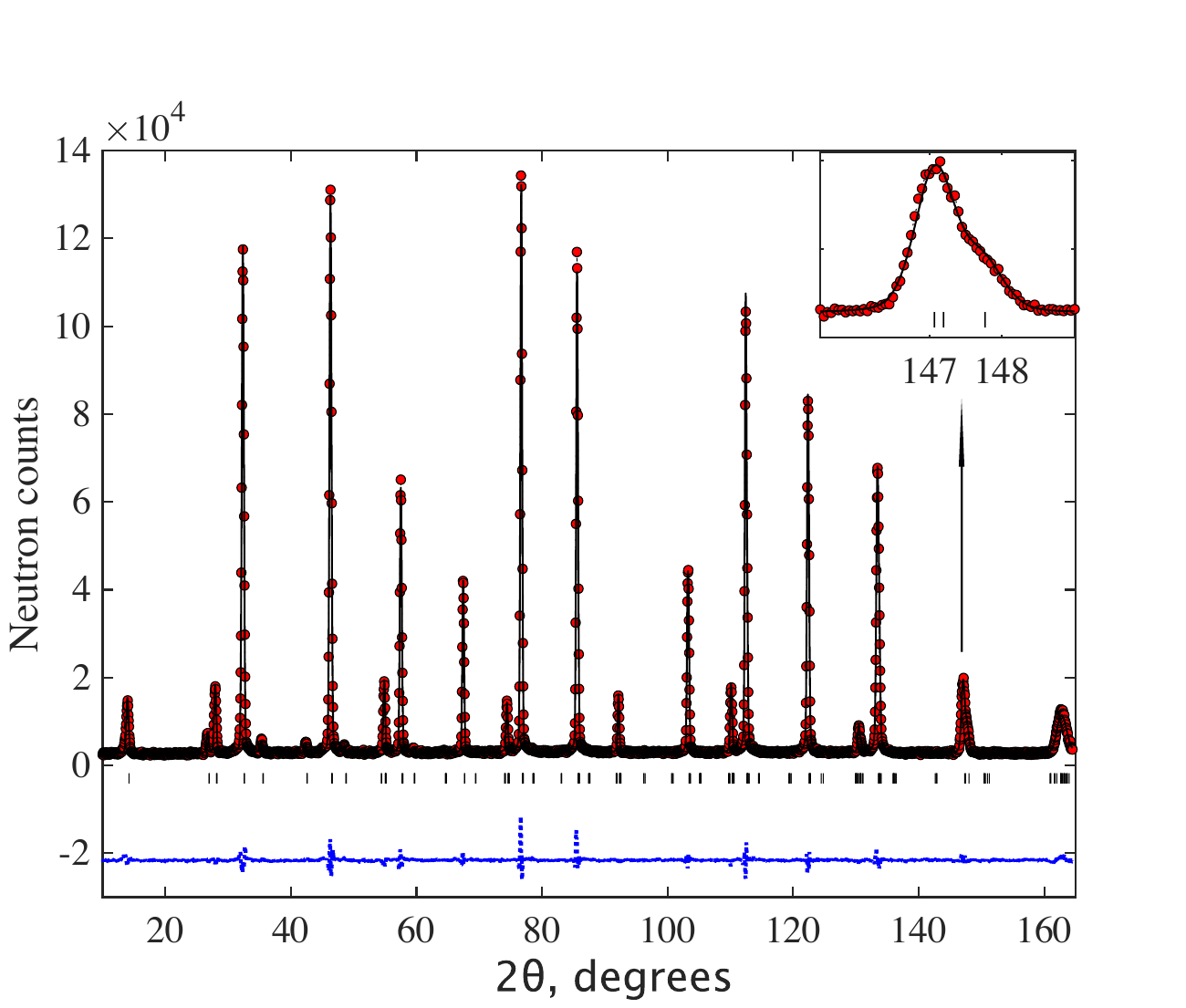}

  \end{center}
\caption{The Rietveld refinement pattern and the difference plot of the neutron diffraction pattern for MnO (top) and NiO (bottom) at $T=2$~K  measured at HRPT with the wavelength $\lambda=1.1545$~\AA. The line is the refinement pattern based on the magnetic model $C_c2/c$ shown in Table~\ref{tab_str}. The row of tics shows the Bragg peak positions. The insets show fragments of the diffraction pattern at high angles. }
\label{dif_MnO}
\end{figure}

The diffraction patterns were collected at a base temperature of T=2~K for all samples and at some selected temperatures below and above the N\'eel temperature $T_N=115$~K for MnO. 
Diffraction pattern for NiO taken at room temperature, about 300~K, is similar to the one at base temperature, because the $T_N=525$~K is significantly higher than room temperature. Figure \ref{dif_MnO} shows the diffraction patterns of MnO and NiO at 2K together with the refinement curves. One can see that the rhombohedral distortion in MnO is very large: the group of five peaks around $2\theta=140^\circ$ is mainly the splitting of the single cubic peak (711)/(551). In \Rm\ metric they are (321), (045), (137), (309), and (2,0,11).  For NiO the rhombohedral splitting is about ten times smaller than for MnO but is  clearly seen as an \Rm\ doublet (404), (0,0,12) from the splitting of the (444) cubic peak, around $2\theta=147^\circ$.  The rhombohedral distortions from the parent cubic metrics can be defined as the ratio $\delta_R=(\sqrt{6}-c/a)/\sqrt{6}$ and amount to $\delta_R\simeq 1.9\cdot 10^{-3}$ and $ 16\cdot 10^{-3}$ for NiO and MnO, respectively, where $a$ and $c$ are rhombohedral lattice constants.

The structure can be further refined in \Cc, leading to additional monoclinic splitting of the above peaks. The fit to the monoclinic group is converged with the metrics shown in Table \ref{tab_str}. We note that all the atoms preserve their special positions in the \Cc symmetry, so there are no extra structure parameters to refine, allowing a convergent fit.
The monoclinic distortion can be refined as well. Despite the fact that the structures can be refined in \Cc, there is no apparent peak splitting and the improvement in the goodness of fit $\chi^2$ is marginal. The refined monoclinic distortion from the rhombohedral metric corresponds to the splitting of the Bragg peaks (shown in the inset of Fig.~\ref{dif_MnO}) $\delta d/d \simeq 3\cdot 10^{-4}$ and $6 \cdot 10^{-4}$ for the NiO doublet around $2\theta=147^\circ$ and MnO around $2\theta=139.2^\circ$, respectively. The metric transformation from  \Rm\ to \Cc\ reads: (-1,-2,0), (-1,0,0), (2/3,4/3,-2/3). This transformation  results in the following monoclinic metric that preserves R-symmetry: $A=\sqrt{3}a$, $B=b$, $C={2\over3}\sqrt{3a^2+c^2}$, $\beta=180-\arccos(a/\sqrt{a^2+c^2/3} ) $, where capital letters stand for the monoclinic metric. This ``ideal'' value of $\beta$, derived by the above formula, is also listed in Table~\ref{tab_str}. We think that the refined monoclinic deviations from the rhombohedral symmetry may suffer from systematic errors related to profile parameters. From the viewpoint of high-resolution neutron diffraction, the crystal metric is indistinguishable from the rhombohedral one. 


\subsection{One-\kI\ magnetic structure}

All magnetic Bragg peaks can be indexed with a single propagation vector \kI\ in \Fm\ settings. More precisely, due to very strong rhombohedral distortions, the propagation vector that fits the experimental peak positions is \kIIR\ in \Rm\ settings, with the basis transformation from the cubic group shown in Table \ref{tab_str}. The one-$k$ rhombohedral magnetic subgroup $R_I\bar{3}c$ has the magnetic ion in the 6b Wyckoff position, which allows a spin component only parallel to \kIIR\ and gives the structure with ferromagnetic $(ab)$ layers with an antiparallel orientation of neighbouring  layers along $c$.  This type of structure does not fit the experiment because it does not explain the first magnetic peak \kIIp\ (in cubic setting) or (003) in the above rhombohedral group.  

\begin{figure}
  \begin{center}
    \includegraphics[width=\figsiz]{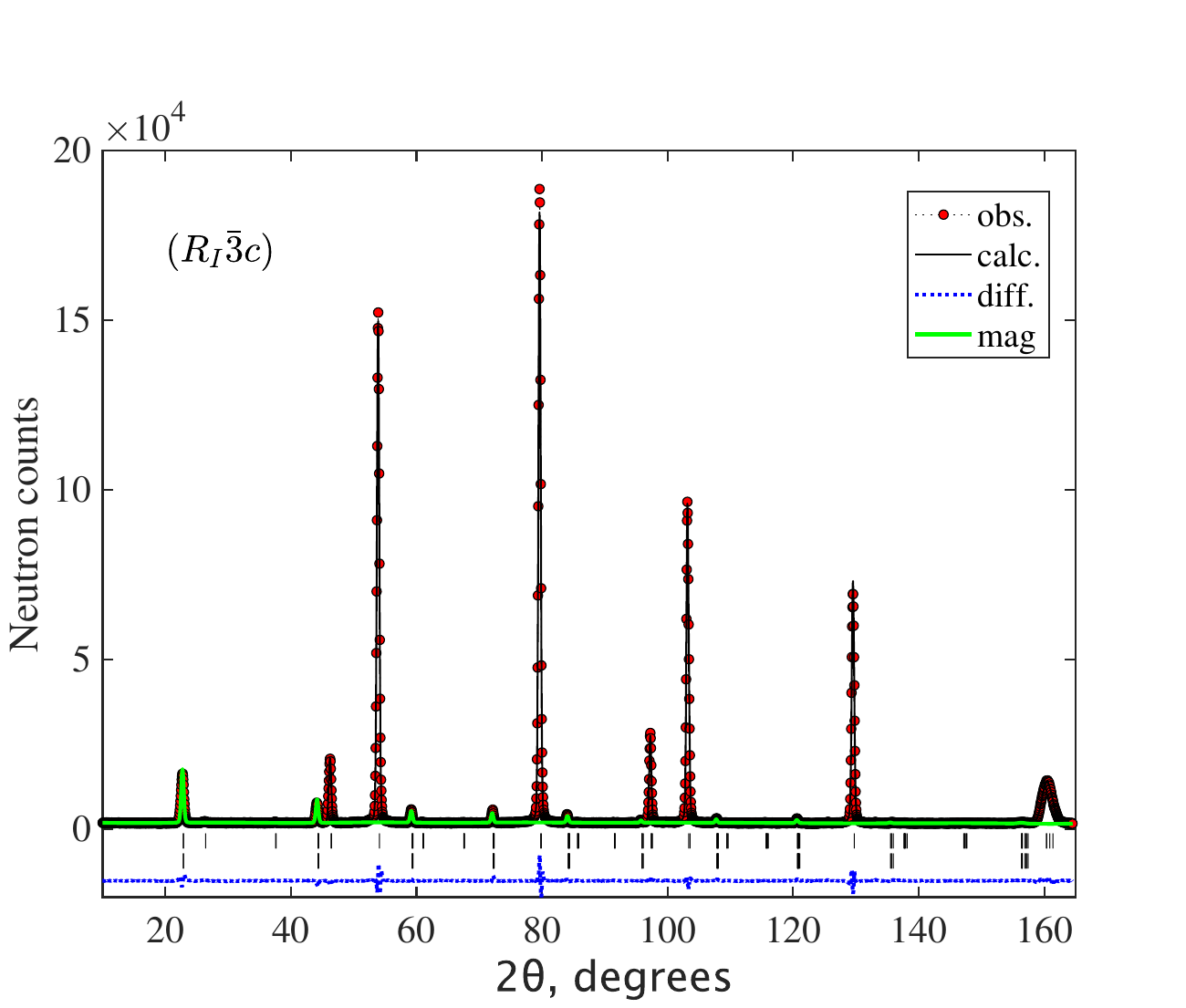}
    \includegraphics[width=\figsiz]{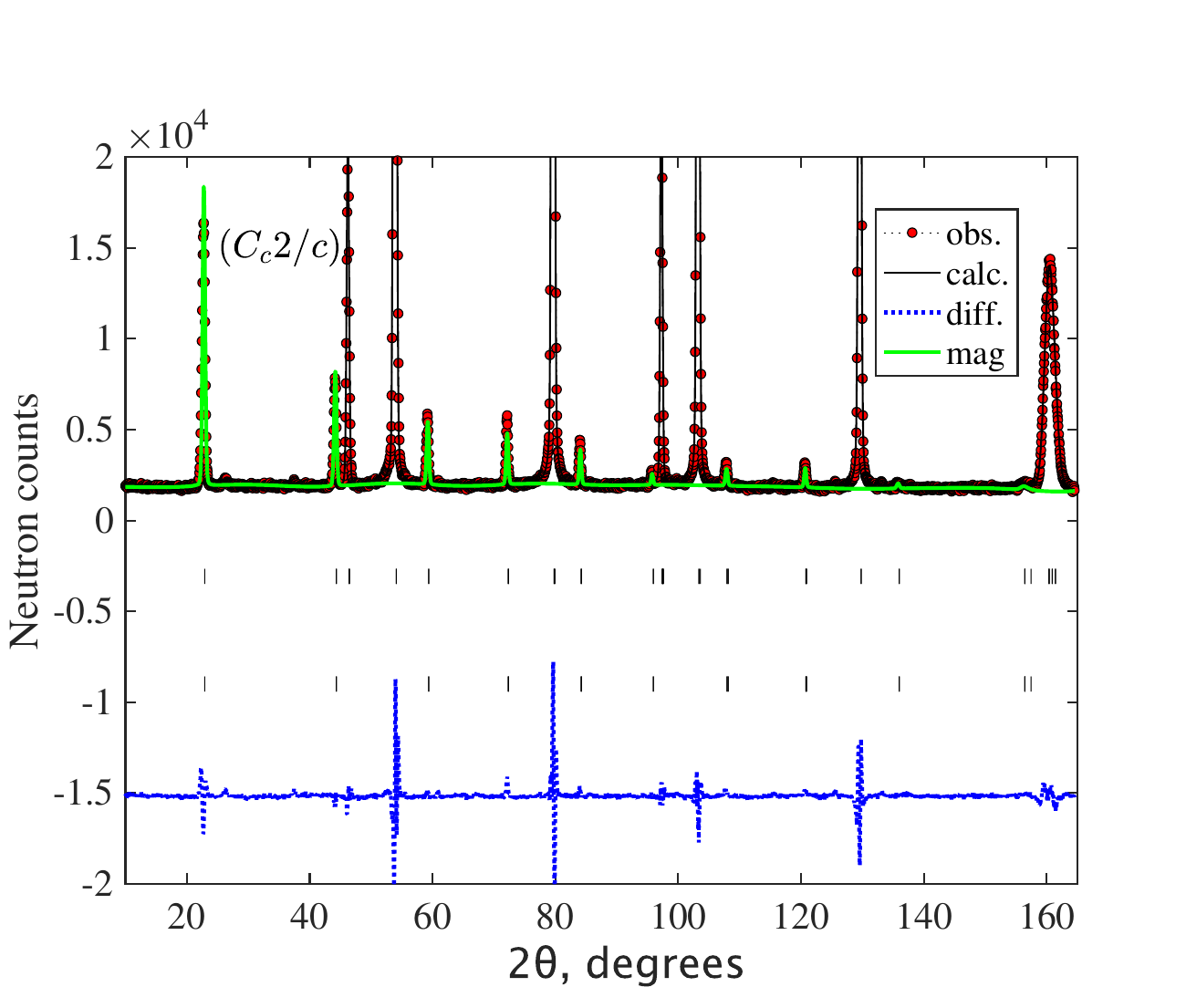}
  \end{center}
\caption{The Rietveld refinement pattern and the difference plot of the neutron diffraction pattern for NiO at $T=2$~K  measured at HRPT with the wavelength $\lambda=1.886$~\AA\ with high statistics. The lines are the refinement patterns based on the magnetic model \RIc\ (top) and $C_c2/c$ (bottom) shown in Table~\ref{tab_str}. The magnetic contributions are  shown by the green lines. The rows of tics show the Bragg peak positions for the crystal (upper) and magnetic (lower) structures. The difference between the observed and calculated intensities is shown by the dotted blue line. Since both refinements are visually identical, we zoom the bottom $C_c2/c$ figure to better show the magnetic contribution.  The minor unindexed peaks are due to $\lambda/2$, as discussed in the text and illustrated in Fig.~\ref{dif_NiO_1p9_C7_l2}.}
\label{dif_NiO_1p9_C7} 
\end{figure}

\begin{figure}
  \begin{center}
    \includegraphics[width=\figsiz]{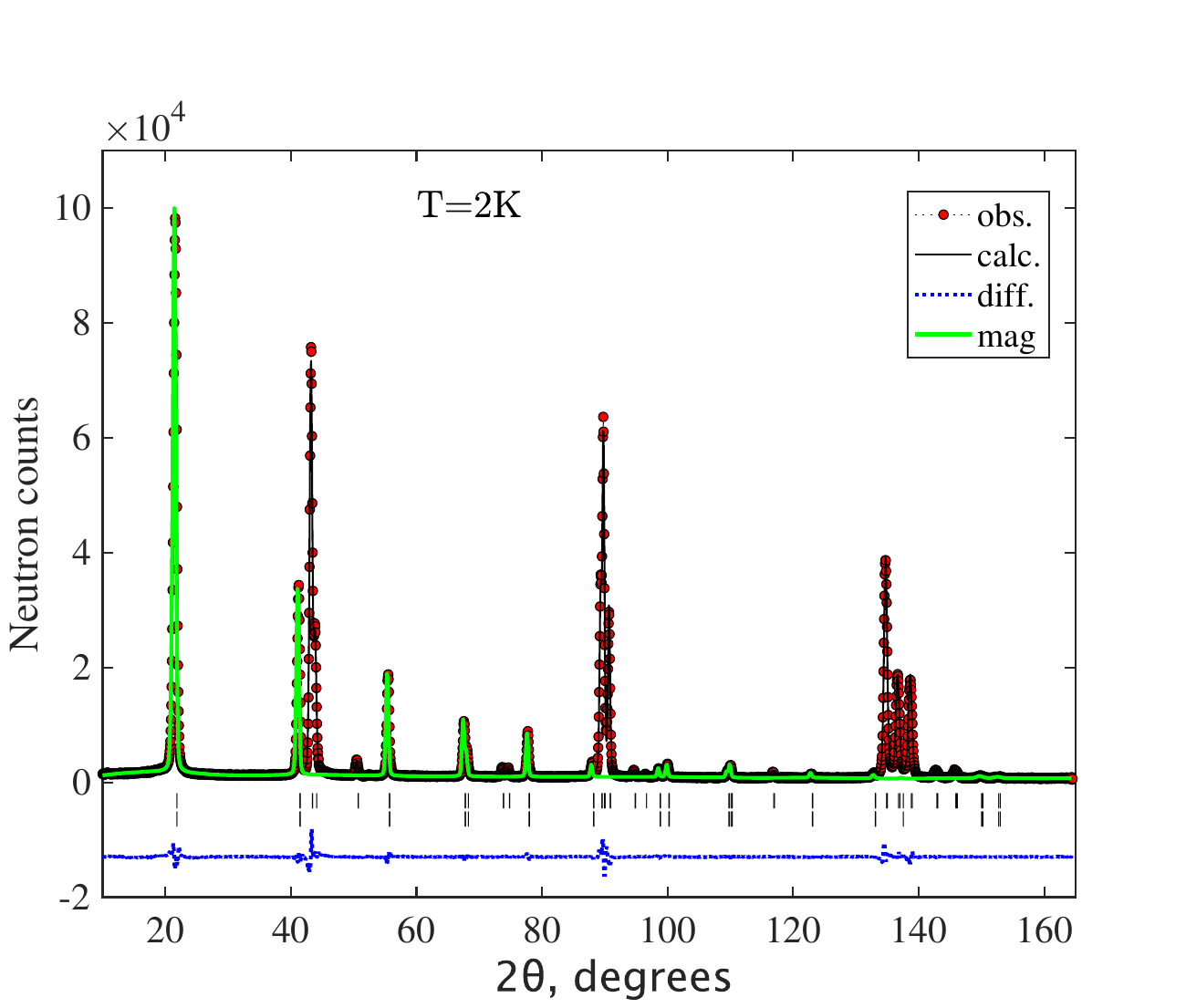}
    \includegraphics[width=\figsiz]{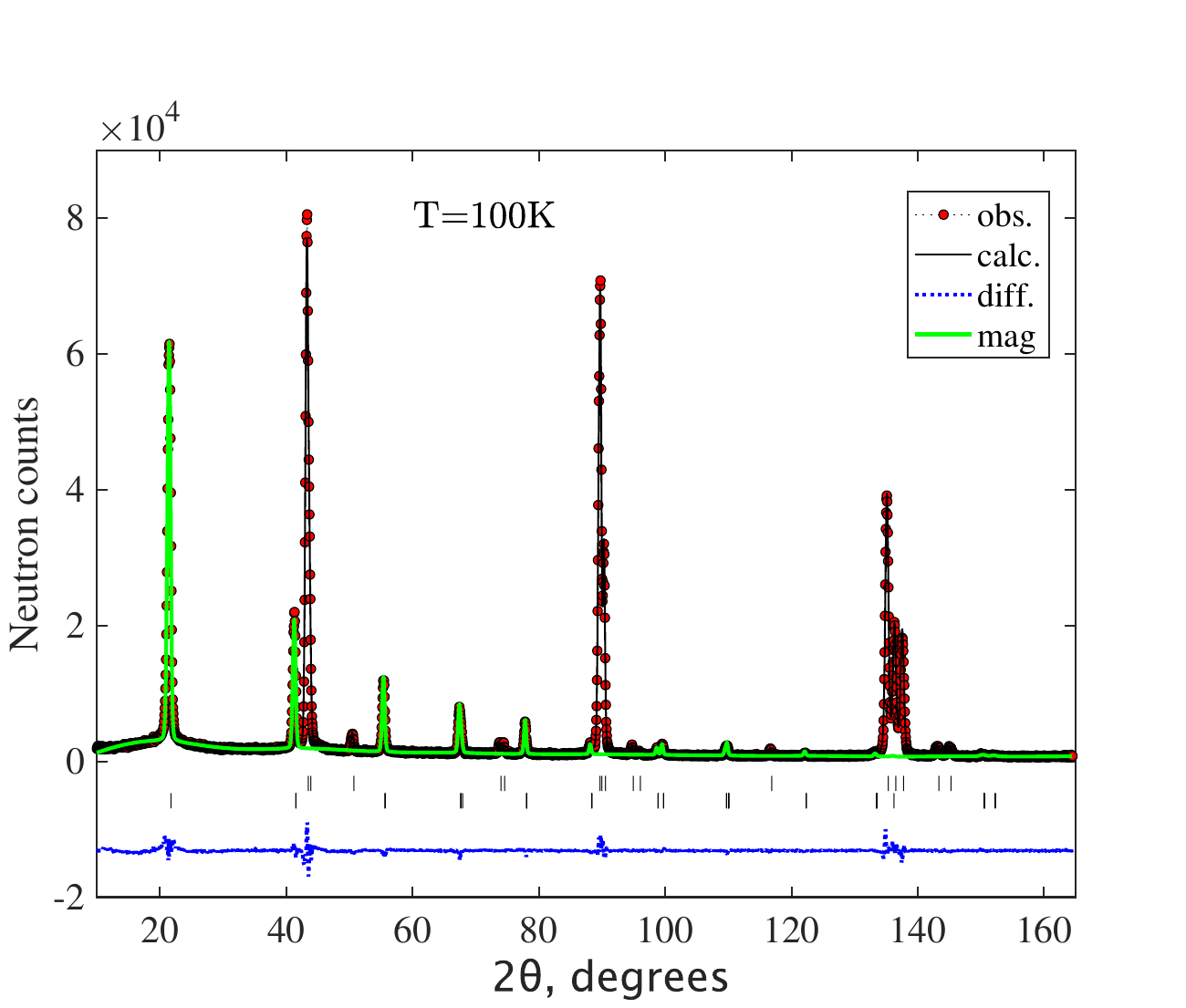}
    \includegraphics[width=\figsiz]{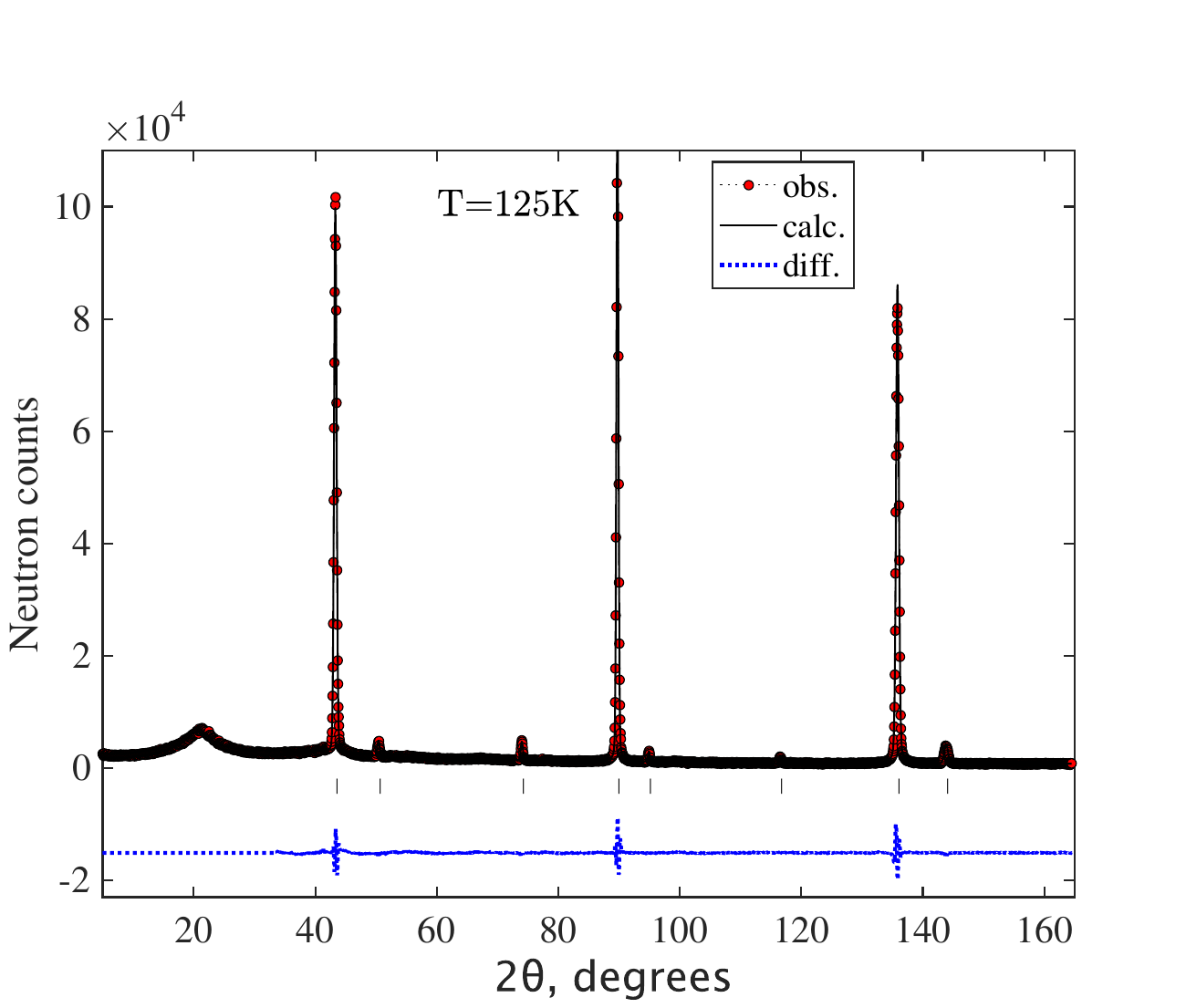}
  \end{center}
\caption{ The Rietveld refinement pattern and the difference plot of the neutron diffraction pattern for MnO at $T=2$~K (top), 100~K (middle), and 125~K (bottom) measured at HRPT with the wavelength $\lambda=1.886$~\AA. The lines are the refinement patterns for MSG $C_c2/c$ below the N\'eel temperature for $T=100$~K and the paramagnetic SG \Fm\ for 125~K.  The magnetic contributions are shown by the green lines. The rows of tics show the Bragg peak positions for the crystal (upper) and magnetic (lower) structures. The difference between the observed and calculated intensities is shown by the dotted blue line. }
\label{MnO_diffraction} 
\end{figure}

\begin{figure}
  \begin{center}
    \includegraphics[width=\count1cm]{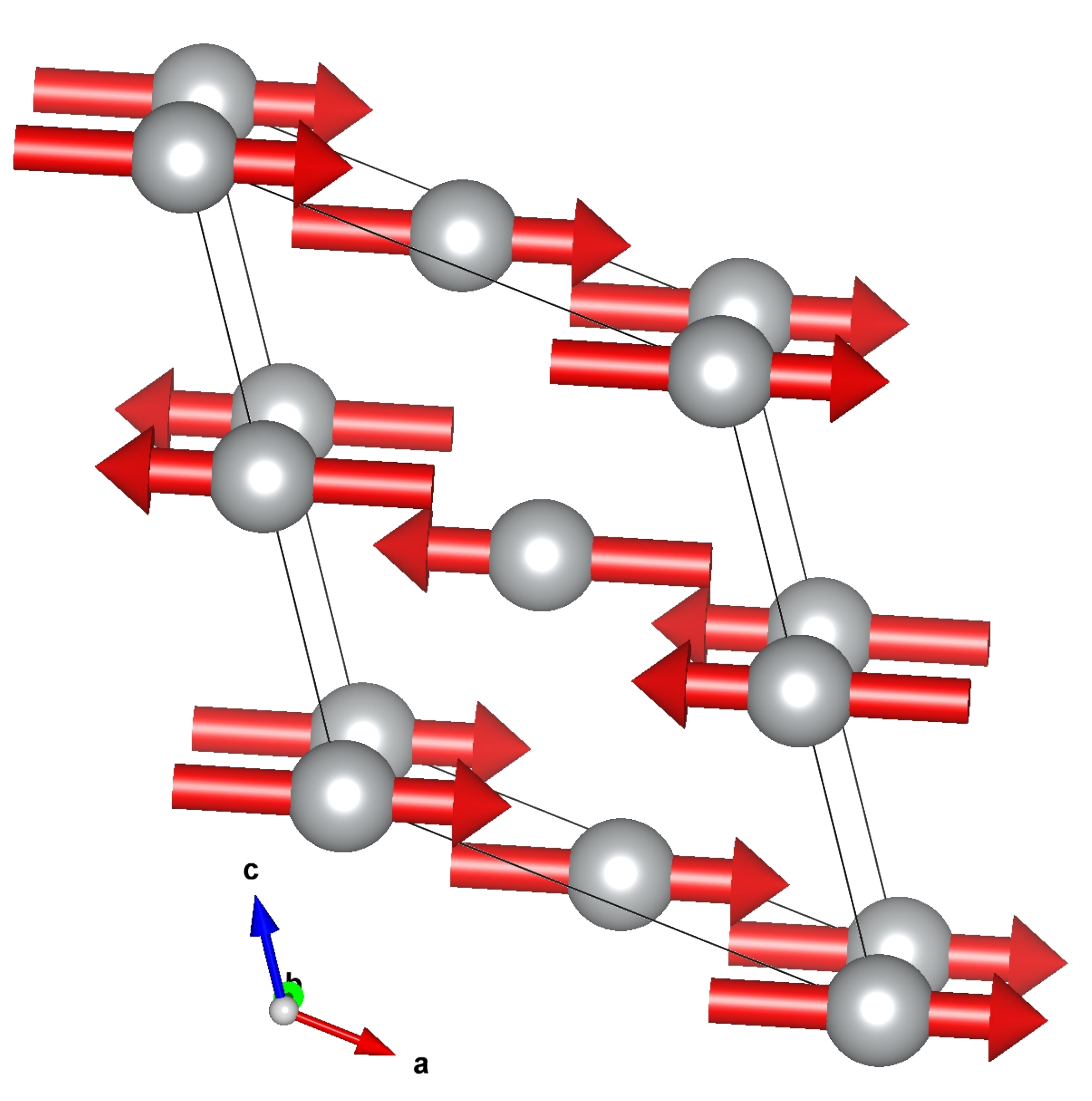}
    \includegraphics[width=\count1cm]{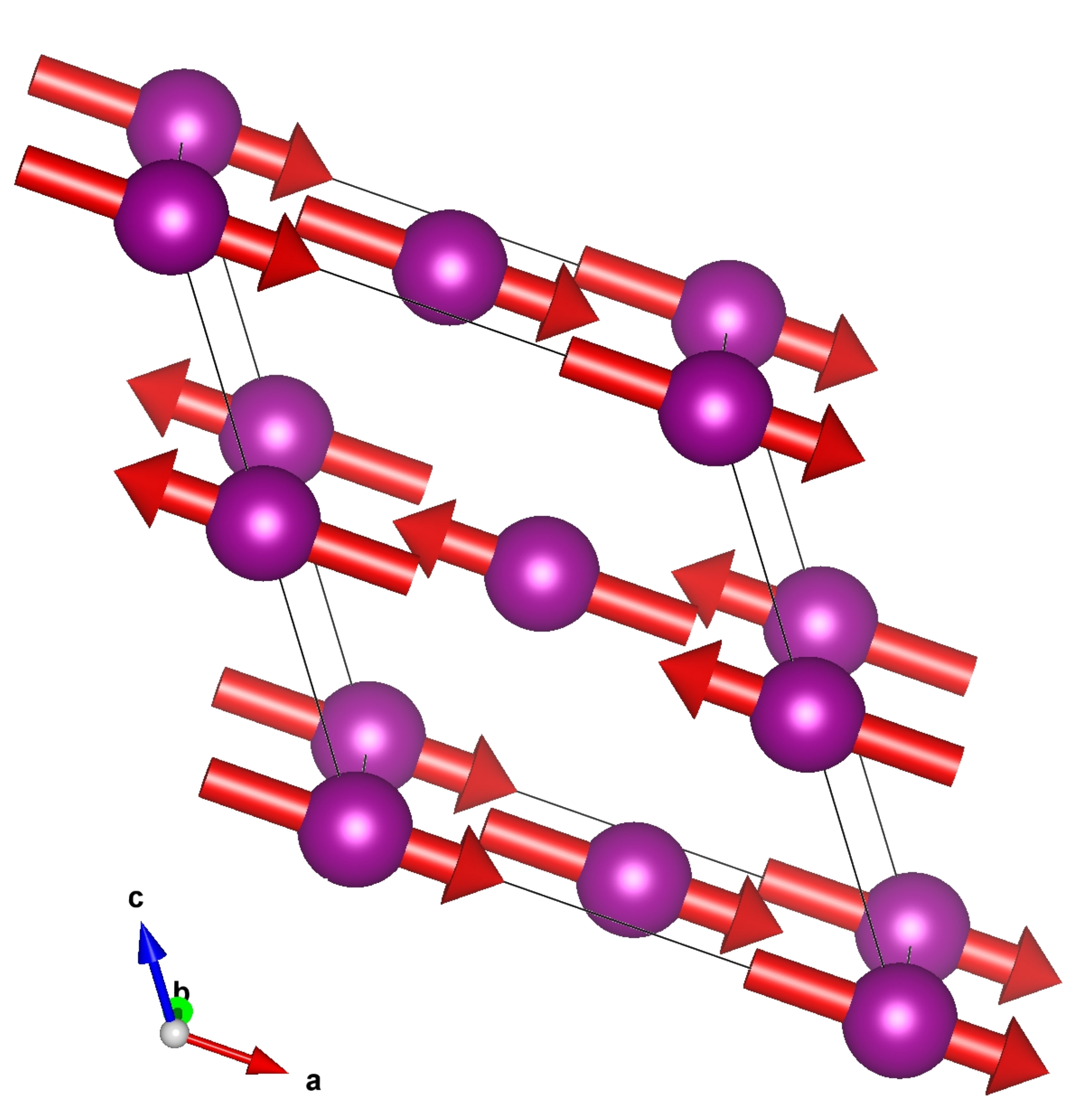}
  \end{center}
\caption{Magnetic structure of NiO (left) and MnO (right) in MSG $C_c2/c$. The details of the magnetic models are given in Table~\ref{tab_str}.}
  \label{MS_C2c_1k}
\end{figure}

The data are very well fitted in the $C_c2/c$ MSG (Fig.~\ref{dif_NiO_1p9_C7}) with the structure parameters shown in Table \ref{tab_str}. The directions of moments along specific monoclinic axes are dictated only by the symmetry of the magnetic space group and are conditional because the crystal metric does not show any explicit deviations from the rhombohedral one, according to our experimental data. There was no convergence of the fit with the $m_z$ component released for MnO, so it was fixed to $m_z=0$. Since the experimental data do not show monoclinic distortions, the absence of a spin component along the \kI\ vector raises the symmetry of MSG to $C_c/2m$, where the allowed spin direction is only $m_y$, which is perpendicular to \kI, similar to $m_x$. The fit in MSG $C_c/2m$ gives the same goodness of fit and the same magnetic moment for MnO.
The description of the magnetic structure for both samples is given in $C_c/2c$ in Table~\ref{tab_str}. The formal description of the magnetic structures in the mcif format is given in the Appendix section~\ref{mcif_app} and also in the supplementary information to this paper~\cite{SM}. The Figures \ref{dif_NiO_1p9_C7} and \ref{MnO_diffraction} show the neutron diffraction patterns together with the fits to the magnetic structure models. For MnO, the figure shows the diffraction patterns below and above the N\'{e}el temperature $T_N=115$~K, where one can see progressive development of the rhombohedral distortions with the temperature lowering  at $2\theta\simeq 90$ and 140~degrees. The presence of the short-range magnetic correlations just above $T_N$ is also well seen as a broad \kIIp\ peak at $2\theta\simeq 20$~degrees. The magnetic structures are illustrated in Figure \ref{MS_C2c_1k}. The magnetic moment sizes are close  to the expected spin-only values of 2 and 5~$\mu_B$ for NiO and MnO, respectively. 

%
The structure description in \Rm, given in Table \ref{tab_str}, does not correspond to any rhombohedral MSG but is equivalent to the magnetic structure with MSG $C_c2/c$, with the crystal metric fixed as explained in section \ref{crys_sec}. In this description, the crystal structure is refined in the \Rm\ space group, whereas the magnetic structure is defined in $P1$ with rhombohedral centering translations ($R1$).  In the rhombohedral metric, the moment components are along and perpendicular to the $c$-axis.  For \nio, there is a spin component parallel to the propagation vector, whereas it is absent for \mno. This type of representation is useful for  comparison with the literature data and the multi-$k$ model considered in the next section.

\subsection{Multi-$k$ magnetic structures in NiO}
\begin{figure}
  \begin{center}
    \includegraphics[width=\count1cm]{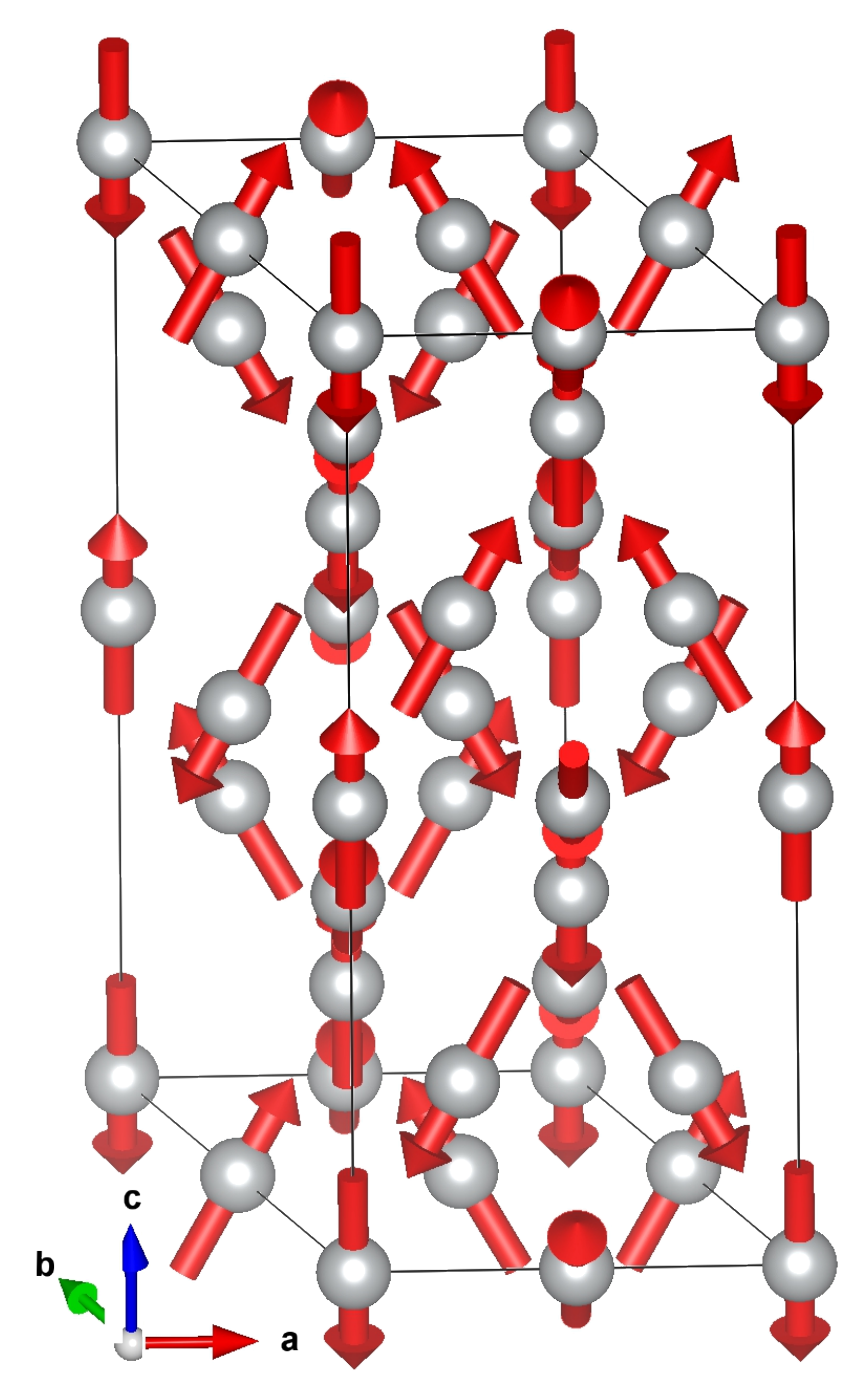}
  \end{center}
\caption{Magnetic structure of NiO multi-$k$ MSG $R_I\bar{3}c$. The details of the magnetic model are given in Table~\ref{tab_str}.}
  \label{MS_RI3_4k}
\end{figure}

Two rhombohedral groups, $R_I\bar{3}c$ (167.108 $R\bar{3}c.1'_c[R\bar{3}m]$(UNI symbol)) and its subgroup $R_I\bar{3}$ (148.20 $R\bar{3}.1'c[R-3]$ ), allow a spin component on both magnetic sites 18e and 6a. 
We performed a simulated annealing (SA) search \cite{kirkpatrick83,Fullprof} of the full diffraction profile for possible solutions in the most general rhombohedral MSG $R_I\bar{3}$ for both NiO and MnO samples. 
The SA search  starts from random values of the free parameters, and we repeated the search more than several hundred times, so we are confident that we have not missed any solutions.
%
For MnO, there is no reasonable multi-$k$ rhombohedral solution in $R_I\bar{3}$ that would be compatible with the 1$k$ structure based on the goodness of fit. For NiO, we have found two structures that give a goodness of fit similar to the one-$k$ structure in the $C_c2/c$ 1$k$ model.  The results of the SA search are shown in Table \ref{tab_str}. 
One of the structures has ``good'' sizes of the magnetic moments, about 2~$\mu$B, which is shown on the first line for the $R_I\bar{3}$ model SA search in Table~\ref{tab_str}. One can see that the components along $x$ and $y$ are related by a factor of two, suggesting the $R_I\bar{3}c$ symmetry. Using these initial values of the parameters, we  performed a standard least squares (LSQ) fit, but the fit did not converge for all parameters released. We attempted a restricted fit under the assumption of the same moment sizes on both Ni sites using spherical coordinates with a fixed ratio of $m_x/m_y=1/2$, but there was no convergence for the spherical $\theta$ angle to the $z$-axis. So, in the final LSQ fit, the angle was fixed to $-35$~degrees (close to the value from the SA-search). The error bars correspond to the $\theta$ variation between $-30$ and $-40$~degrees. The formal description of the structure is given in Appendix~\ref{mcif_app}. The illustration of the fit quality can be seen in Fig.~\ref{dif_NiO_1p9_C7}. The magnetic structure in this multi-$k$ model in MSG $R_I\bar{3}c$ is shown in Fig.~\ref{MS_RI3_4k}. The magnetic structure description with respect to the atomic positions and the crystal metrics is the same in both MSG  $R_I\bar{3}c$ and its subgroup $R_I\bar{3}$, provided that we correlate $m_y=2m_x$ in the latter. In the Appendix~\ref{mcif_app}, we give the description of the structure in $R_I\bar{3}$. For completeness, we also provide an mcif file with the equivalent description of this structure in MSG $R_I\bar{3}c$ in Supplementary Materials~\cite{SM}.

\section{Discussion}

The magnetic structure in MnO can be described using two different one-$k$ MSGs, $C_c2/c$ and $C_c2/m$, which give different magnetic moment directions with respect to the same monoclinic metrics in these groups. Since our diffraction data do not show significant deviations from the rhombohedral crystal metrics, we cannot unambiguously assign the MSG symmetry.  We note that the above MSGs are not group-subgroup related. Therefore, we use the less restricted monoclinic MSG $C_c2/c$ for the magnetic structure description for both samples. Furthermore, as we have shown, our data for both MnO and NiO samples are not sensitive to the orientation of the magnetic moment in the plane perpendicular to the rhombohedral $c$-axis in $R\bar{3}m$ SG (direction of \kI=\kII\ in parent SG \Fm) --- we can only determine the components of the magnetic moment along and perpendicular to \kI.
%

\begin{figure}
  \begin{center}
    \includegraphics[width=\figsiz]{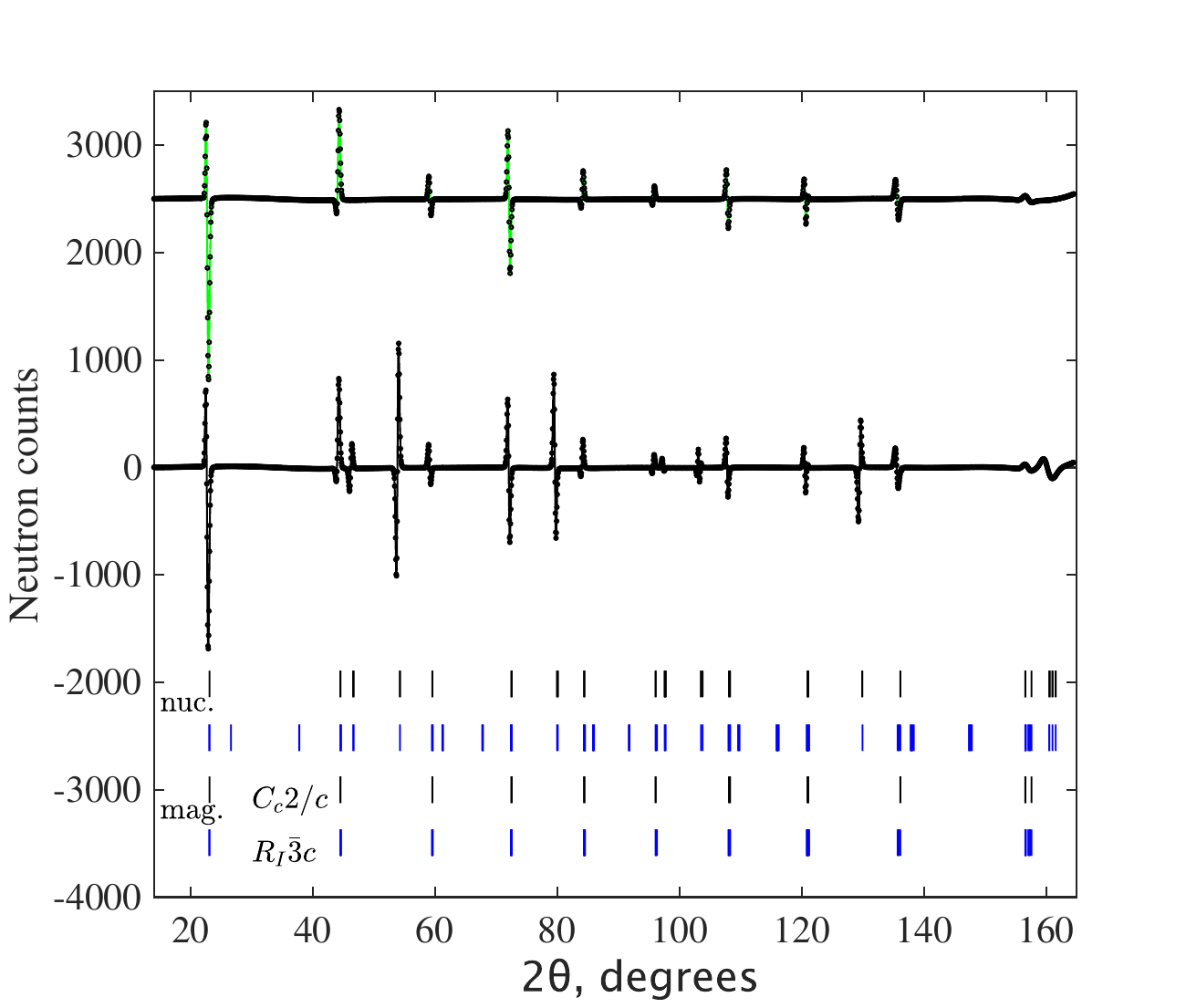}
  \end{center}
\caption{The difference between the calculated neutron diffraction patterns shown in Fig.~\ref{dif_NiO_1p9_C7} for the models \RIc\ and $C_c2/c$ with refinement parameters shown in Table~\ref{tab_str} for NiO at $T=2$~K. The line around the zero level is the overall difference between the calculated patterns. The line shifted up to the level 2500 represents the difference in pure magnetic contributions between the two models. The rows of tics show the Bragg peak positions for the crystal (upper) and magnetic (lower) structures.}
\label{dif_dif_Cc_RI} 
\end{figure}

The neutron diffraction patterns allow two very different magnetic structure models for NiO, which fit the data equally well. The directions of the magnetic moments in the monoclinic one-$k$ structure $C_c2/c$ and the rhombohedral 4$k$ structure $R_I\bar{3}c$ are practically orthogonal to each other. In the one-$k$ structure, the Ni moments form ferromagnetic (FM) layers with the moment direction approximately perpendicular to the propagation vector \kI\ and stacked antiferromagnetically (AFM) along the \kI. In the multi-$k$ magnetic structure, the moments of the neighbouring Ni atoms are approximately AFM coupled. These two structures are completely incompatible with each other, and we think that the 4$k$ structure $R_I\bar{3}c$ is a false solution and is only possible because of the very small rhombohedral distortions in NiO, as well as the small moment value. We cannot formally disregard this magnetic structure because it gives the same goodness of fit, but we would like to show the reasons for excluding this model. Figure~\ref{dif_dif_Cc_RI} shows a comparison between the models. The lines show the difference between calculated diffraction profiles for the models \RIc\ and $C_c2/c$. The allowed magnetic Bragg reflections are localized at approximately the same two-theta values, but due to rhombohedral distortion, all peaks are split in the larger unit cell in \RIc\ model. For instance, the first magnetic peak at 22.5~degrees, which is a single peak (001) in $C_c2/c$, becomes a doublet in \RIc\ (101)/(003) with (101) at a slightly lower two-theta. Thus, the peak (003) has zero intensity in \RIc, but all calculated intensity belongs to the (101) peak. Conversely, the single peak (001)  in $C_c2/c$, metrically corresponding to (003), describes all the experimental intensity. This repartitioning of the peak intensities results in the up-down shape of the calculated intensity difference clearly seen at all magnetic peak positions in Fig.~\ref{dif_dif_Cc_RI}. 
Although the models in Figures~\ref{MS_C2c_1k} and \ref{MS_RI3_4k} may be indistinguishable by neutron diffraction, they almost certainly have very different total energies. {\it Ab initio} calculations can and should be used to select the correct model in cases like NiO.

\begin{figure}
  \begin{center}
    \includegraphics[width=\figsiz]{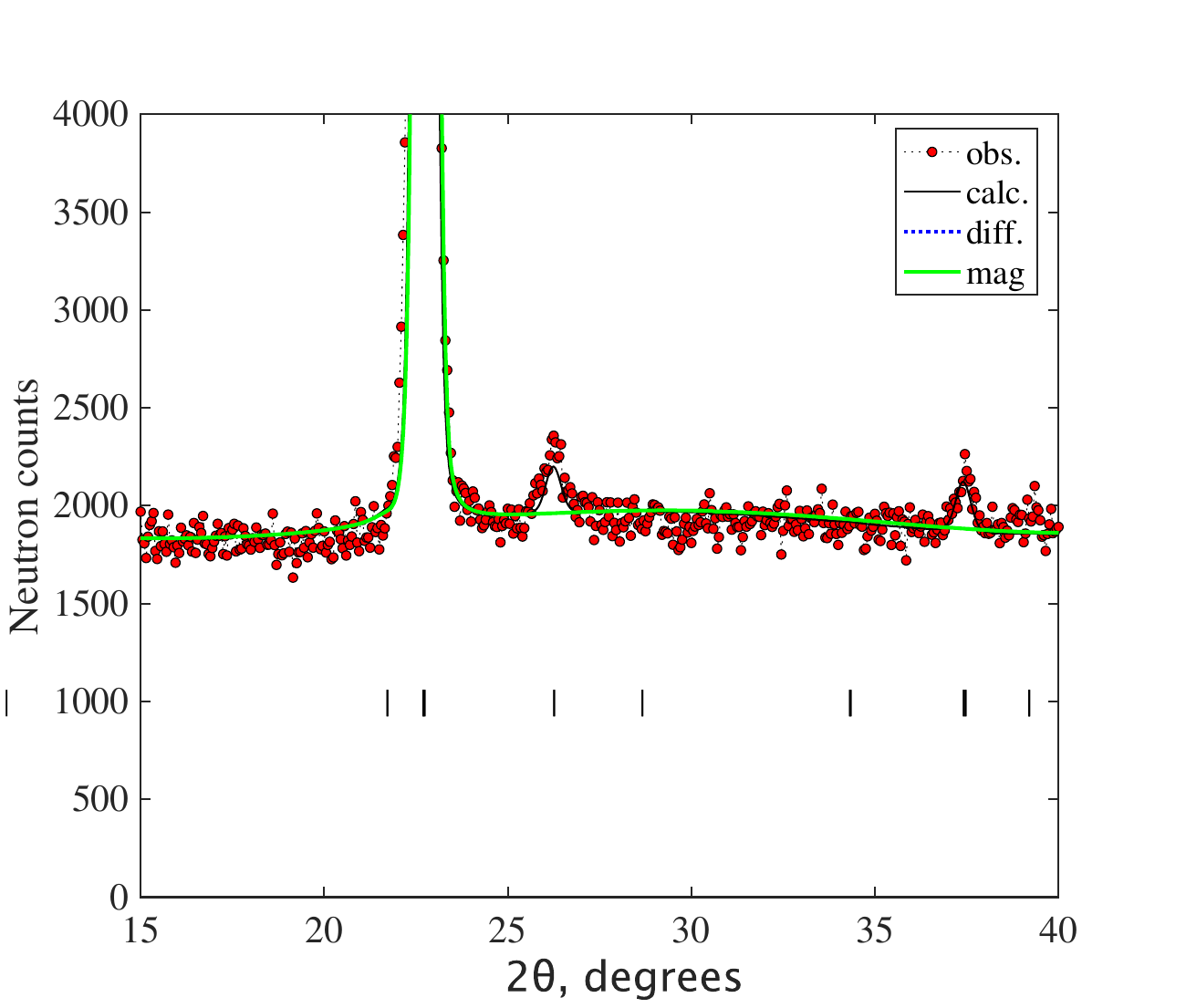}
  \end{center}
\caption{A portion of the Rietveld refinement pattern for NiO at $T=2$~K  measured at HRPT with the wavelength $\lambda=1.886$~\AA\ with high statistics, demonstrating the presence of a very small but visible contribution (about 0.05\%) of $\lambda/2$ at two-theta values of approximately 26 and 37 degrees, corresponding to the strongest Bragg peaks (111) and (311) in cubic metric. The line is the refinement pattern for the  $C_c2/c$ model shown in Table~\ref{tab_str}. The magnetic contribution is shown by the green line. The row of tics shows the Bragg peak positions.}
\label{dif_NiO_1p9_C7_l2}
\end{figure}

The presence of the extra Bragg peaks in the larger unit cell of \RIc\ could potentially prove the multi-$k$ magnetic structure, although these peaks have only nuclear contributions in \RIc. Close inspection of the diffraction pattern shows that a few peaks are indeed present (Fig.~\ref{dif_NiO_1p9_C7_l2}), but they originate from a very small admixture of the $\lambda/2$ component in the neutron beam, and their intensities are ideally refined in one-$k$ $C_c2/c$ model. Ironically, we did not realize the presence of 0.05\% of $\lambda/2$ in the neutron beam, which theoretically should not be present for the (511) reflection of the Ge-monochromator used for $\lambda=1.886$~\AA. Thus, the presence of these extra peaks was initially attributed to the multi-$k$ structure, and we attempted to find the solutions in all possible multi-$k$ structures shown in the subgroup tree (Fig.~\ref{k_graph}).

In MnO, due to significantly larger rhombohedral distortions, one can clearly see that the intensity distribution over the doublets like (101)/(003) definitively allows us to reject the rhombohedral model. We believe that the exchange interactions in both compounds are similar, and also on these grounds, the monoclinic model is the only acceptable solution for the magnetic structure in NiO.  
An interesting previous study~\cite{PhysRevLett.96.047209} reports that the monoclinic, collinear structure of MnO is favored when considering the total (Bragg + diffuse) neutron scattering, and it also resolves the direction of the moment in the (111) planes. The solution proposed is similar to the one we have found, but with a different monoclinic symmetry, $C2$. This group cannot be a magnetic subgroup of the parent gray MSG, but it can, of course, be modeled using the lowest MSG $P_S\bar{1}$.

We would like to note that distinguishing between multi-$k$ and one-$k$ structures from diffraction experiments is difficult, if not impossible, unless there is a magnetostructural distortion associated with the different symmetries of the multi-$k$ vs. one-$k$ cases. The point is that different $k$-vectors from the propagation vector star do not interfere with each other, and their partial contributions are simply added to the Bragg peaks located at the same $Q$-value in powder diffraction. The one-$k$ structure contributes to the Bragg peaks located at the same $Q$, as in the multi-$k$ case, and we think that in most cases one can find one-$k$ and multi-$k$ models that provide the same integral intensity at the same $Q$. However, this is not always possible if we keep the same magnetic space group symmetry. For instance, in our case, the 1$k$ and 4$k$  magnetic structures with very different arrangements of the magnetic moments in the same MSG $R\bar{3}c$ ($mL2^+$) result in different magnetic Bragg peak intensities, and the 1$k$ structure can be definitely rejected. However, it is true that one can find the one-$k$ structure in monoclinic MSG $C_c2/c$, which gives the same powder diffraction intensities in the absence of crystal metric distortions.
The situation is not much easier in single-crystal diffraction where, instead of powder averaging, we always have magnetic domains for the one-$k$ structure that mimic the multi-$k$ case.

In some cases, one can distinguish between the models using single-crystal neutron diffraction in an applied magnetic field because multi-$k$ and one-$k$ structures will be modified differently (see, for instance,~\cite{CeAlGe}). Alternatively, one could potentially affect the population of the magnetic domains by applying uniaxial pressure, thus identifying the one-$k$ model. Additionally, there is an example of inelastic neutron scattering on powder samples where the authors claim that they can distinguish multi-$q$ from single-$q$ states~\cite{nature_INS} from the analysis of the magnetic excitations in linear spin-wave theory.


\section{Summary}

We have performed high-resolution powder neutron diffraction experiments with MnO and NiO monooxides with N\'{e}el temperatures $T_N=$~115~K and 525~K, respectively, in order to refine their crystal and magnetic structures. The structures below $T_N$ are generated by a propagation vector \kI=\kII\ in the parent paramagnetic cubic space group (SG) \Fm. The rhombohedral distortions from the parent cubic metrics are well seen in our diffraction data and were refined in the $R\bar{3}m$ SG, amounting to $1.9\cdot 10^{-3}$ and $ 16\cdot 10^{-3}$ for NiO and MnO, respectively. In accordance with the rhombohedral symmetry, there are eight possible magnetic space subgroups (MSG) for the one-$k$ and multi-$k$  structures. Among them, there are three most symmetric MSGs: \RIc, $C_c2/m$ and \Cc. The refined monoclinic distortions from the rhombohedral metric are very small, about a few $10^{-4}$, and can be disregarded,  implying  that the orientation of the magnetic moment in the plane perpendicular to the rhombohedral $c$-axis cannot be inferred from our data.

We have found that the best antiferromagnetic (AFM) structure, which fits both MnO and NiO, is the 1$k$ structure with MSG $C_c2/c$ (BNS 15.90, UNI symbol $C2/c.1'_c[C2/m]$). In this structure, the moments form ferromagnetic (FM) layers with the moment direction approximately perpendicular to the propagation vector \kI\ and stacked AFM along the \kI\ with magnetic moment sizes $m=1.94$ and 4.61~$\mu$B for NiO and MnO, respectively. The same magnetic configuration can be described equally well in  the MSG $C_c2/m$ for MnO. For NiO, the multi-$k$ rhombohedral structure in MSG \RIc, which  is very different from the \Cc-structure, can equally well fit our diffraction data with a similar value of $m$.  However, we conclude that this is a false solution and is only possible because of the very small rhombohedral distortions and the small magnetic moment value in NiO. 

In conclusion, we would like to note that this is not an uncommon case where the crystal distortions due to magnetoelastic coupling below the magnetic transition are small or even undetectable in diffraction experiments. Since a paradigm such as ''one propagation vector at the magnetic transition is enough'' is still strong in the neutron scattering community, some high symmetry multi-$k$ magnetic structures are not even considered as possible structure solutions, which could, however,  fit the experimental data equally well.

\section{A{\lowercase{cknowledgements}}}

The work was performed at the Swiss neutron spallation source SINQ at Paul Scherrer Institute.

\bibliographystyle{iucr}
\bibliography{NiMnO_pomjakushin_arxiv_arx.bbl}

%
%

\begin{table} 
\caption{The crystal and magnetic structure parameters in NiO and MnO at T=2K for the 1$k$ monoclinic MSG 
$C_c2/c$ (BNS no. 15.90, UNI $C2/c.1'_c[C2/m]$) and the multi-$k$ $R_I\bar{3}$ (BNS no. 148.20, UNI $R\bar{3}.1'_c[R\bar{3}]$) structure for NiO. The conventional reliability R-factors~\cite{Fullprof}  are also given after the SG symbol for the fits with $\lambda=1.155$~\AA. The reliability factors on the separate lines labeled  $R_{p}$, $R_{wp}$, $R_{exp}$, and $\chi^2$ are for the fits with $\lambda=1.886$~\AA, aimed at the magnetic structure details. The fits in the rhombohedral space group $R\bar{3}m$ with the propagation vector \kIIR\ for the magnetic structures give an equivalent description of structures. The metric transformation from cubic to monoclinic cells reads: (1/2,1/2,-1), (1/2,-1/2,0), (-1,-1,0)  and to 1$k$-rhombohedral reads: (-1/2,1/2,0), (0,-1/2,1/2), (1,1,1). The monoclinic lattice constants derived from the rhombohedral metric are shown after the comma. Ni and O occupy 3a (000) and 3b $(00{1\over2})$ Wyckoff positions in space group $R\bar{3}m$, respectively. 
$m$ is the size of the magnetic moment, $m_x$, $m_y$, and $m_z$ are magnetic moment components in the respective MSGs. 
The magnetic moment has both components perpendicular and parallel to the propagation vector.  In the magnetic space group $C_c2/c$, Ni and O are in 4c $(000 | m_x,0,m_z)$ and 4b $(0{1\over2}{1\over4})$ positions.
The moment directions in monoclinic groups along $x$, $y$, and $z$ axes correspond to $({1\over2},{1\over2},-1)$, $({1},-{1},0)$ and (-1,-1,0) in the cubic metric, respectively.   The Wyckoff positions in the multi-$k$ $R_I\bar{3}$ MSG are 18e (1/2,0,0) and 6a (0,0,0) for Ni1 and Ni2 and 18d (1/2,0,1/4) and 6b (0,0,1/4) for O1 and O2, respectively. Here, $m$ is the size of the magnetic moment, $\theta$ and $\phi$ are spherical angles with $c$ and $a$ axes, respectively, in MSGs $R_I\bar{3}$. 
}

\label{tab_str}

\begin{center} 
\begin{tabular}{l l l} 
& NiO  & MnO \\
 \hline 
$R\bar{3}m$ $\chi^2$ & 8.94, 9.59,  5.84, 2.70 & 8.71, 9.87, 7.18, 1.89\\
$a$, \AA  &   2.95006(3)  & 3.14980(7)\\
 $c$, \AA  &  7.21235(9) & 7.59158(16)\\
 $m_x m_z m$, $\mu_B$
                  &1.840(9)   0.63(3)    1.943(9) & 4.75(1) 0 4.75(1)\\ 
 \hline 
$C_c2/c$  $\chi^2$ & 8.84, 9.01, 5.87, 2.36& 8.46, 9.54, 7.14, 1.79\\
$a$, \AA  &  5.10263(8),   5.10967  & 5.45583(15), 5.45657\\
 $b$, \AA  &  2.95045(6),  2.95007  & 3.15006(9),  3.15035\\
 $c$, \AA  &  5.89158(11), 5.89263 & 6.22984(15), 6.233451\\
 $\beta$, deg.  &  125.1722(9), 125.3159 & 125.6659(13),  125.7027\\
 $m_x\,  m_z\, m$,\, $\mu_B$  
                     &  2.28(2)  0.77(4)  1.946( 9) &  4.61(1) 0 4.61(1)\\
$R_{p},$ $R_{wp}$, $R_{exp}$, $\chi^2$ &  3.67, 5.10, 2.31, 4.86 &  4.51, 6.04, 3.18, 3.61\\

\hline
\multicolumn{3}{c}{SA search for NiO in MSG $R_I\bar{3}$}\\
magnetic moments          & Ni1 (18e) & Ni2 (6a)\\
$m_x m_y m_z m$ , $\mu_B$ &  0.5582   1.1093  -1.6018 1.8678  &0   0   2.0484 2.0484 \\
& -0.5386  -1.1026  0.2224 0.9805   &0   0   -3.4317 3.4317  \\ 
R-factors               &   3.65, 5.27, 2.31, 5.19  &  \\

\multicolumn{3}{c}{LSQ fit in MSG $R_I\bar{3}$}\\
$a,\, c$ \AA&  \multicolumn{2}{c}{5.90046(6), 14.42656(18)} \\ 

$m, \,\mu_B\, \phi, \theta$, deg.&  1.976(9) 90 35(5)       &1.977(9) 0 180 \\ 

$m_x\, m_y\, m_z\, \mu_B$& -0.65(8) -1.31(17) 1.6(1)&-1.977(9) 0 0 \\

$R_{p},$ $R_{wp}$, $R_{exp}$, $\chi^2$    &  \multicolumn{2}{c}{3.72, 5.37, 2.31, 5.38}  \\

 \end{tabular} \end{center}
\end{table}
 
\count1=\figsizcm
\multiply\count1 by 3
\divide\count1 by 5

\clearpage
\section{Appendix: Supplementary Material. Magnetic crystallographic information}
\label{mcif_app}  

Three magnetic crystallographic information files (mcif) are given below.

\renewcommand{\baselinestretch}{0.85}
\begin{verbatim}

#############################################
#
# NiO_P7_C_c2c.mcif
#
#############################################
# NiO 
# One-k magnetic structure in MSG 15.90 C_c2/c, mL3+, OPD P7
_transition_temperature     525
_experiment_temperature     1.8

loop_
_irrep_id
_irrep_dimension
_irrep_small_dimension
_irrep_direction_type
_irrep_action
_irrep_modes_number
_irrep_presence
mL3+  2  2  special  primary   1 .
mL2+  1  1  .        secondary 1  yes

_parent_space_group.name_H-M_alt  "Fm-3m"
_parent_space_group.IT_number      225
_parent_space_group.transform_Pp_abc  "a,b,c;0,0,0"

loop_
_parent_propagation_vector.id
_parent_propagation_vector.kxkykz
k1 [1/2 1/2 1/2]

_parent_space_group.child_transform_Pp_abc  "a/2+b/2-c,a/2-b/2,-a-b;0,0,0"
_space_group_magn.transform_BNS_Pp_abc  "a,b,c;0,0,0"


_cell_length_a     5.10964
_cell_length_b     2.95005
_cell_length_c     5.89292
_cell_angle_alpha  90.00000
_cell_angle_beta   125.31400
_cell_angle_gamma  90.00000
_cell_volume       72.48338

_space_group_magn.number_BNS "15.90"
_space_group_magn.name_UNI "C2/c.1'_c[C2/m]"
_space_group_magn.name_BNS "C_c2/c"
_space_group_magn.number_OG "12.8.73"
_space_group_magn.name_OG "C_2c2/m'"
_space_group_magn.point_group_number_Litvin "5.2.13"
_space_group_magn.point_group_name_UNI "2/m.1'"

loop_
_space_group_magn_transforms.id
_space_group_magn_transforms.Pp_abc
_space_group_magn_transforms.source
1 a,b,c;0,0,0 "BNS"
2 a,b,1/2c;0,1/2,-1/2 "OG"

loop_
_space_group_symop_magn_operation.id
_space_group_symop_magn_operation.xyz
1 x,y,z,+1 
2 -x,y,-z+1/2,+1 
3 -x,-y,-z,+1 
4 x,-y,z+1/2,+1 

loop_
_space_group_symop_magn_centering.id
_space_group_symop_magn_centering.xyz
1 x,y,z,+1 
2 x,y,z+1/2,-1 
3 x+1/2,y+1/2,z,+1 
4 x+1/2,y+1/2,z+1/2,-1 

loop_
_atom_site_label
_atom_site_type_symbol
_atom_site_symmetry_multiplicity
_atom_site_Wyckoff_symbol
_atom_site_fract_x
_atom_site_fract_y
_atom_site_fract_z
_atom_site_occupancy
_atom_site_fract_symmform
_atom_site_U_iso_or_equiv
_atom_site_adp_type
Ni1_1 Ni   4 c  0.00000  0.00000  0.00000  1.00000 0,0,0 0.00031(14)   Uiso
O1_1  O    4 b  0.00000  0.50000  0.75000  1.00000 0,0,0 0.00132(19)   Uiso

loop_
_atom_site_moment.label
_atom_site_moment.crystalaxis_x
_atom_site_moment.crystalaxis_y
_atom_site_moment.crystalaxis_z
_atom_site_moment.symmform
Ni1_1  2.283(18)  0.00000  0.76(4) Mx,0,Mz 

#############################################
#
# MnO_P7_C_c2_c.mcif
#
#############################################
# MnO 
# One-k magnetic structure in MSG 15.90 C_c2/c, mL3+, OPD P7
_transition_temperature     115
_experiment_temperature     1.8

loop_
_irrep_id
_irrep_dimension
_irrep_small_dimension
_irrep_direction_type
_irrep_action
_irrep_modes_number
_irrep_presence
mL3+  2  2  special  primary   1 .
mL2+  1  1  .        secondary 1  no

_parent_space_group.name_H-M_alt  "Fm-3m"
_parent_space_group.IT_number      225
_parent_space_group.transform_Pp_abc  "a,b,c;0,0,0"

loop_
_parent_propagation_vector.id
_parent_propagation_vector.kxkykz
k1 [1/2 1/2 1/2]

_parent_space_group.child_transform_Pp_abc  "a/2+b/2-c,a/2-b/2,-a-b;0,0,0"
_space_group_magn.transform_BNS_Pp_abc  "a,b,c;0,0,0"


_cell_length_a     5.45706
_cell_length_b     3.15147
_cell_length_c     6.23240
_cell_angle_alpha  90.00000
_cell_angle_beta   125.66770
_cell_angle_gamma  90.00000
_cell_volume       87.07706

_space_group_magn.number_BNS "15.90"
_space_group_magn.name_UNI "C2/c.1'_c[C2/m]"
_space_group_magn.name_BNS "C_c2/c"
_space_group_magn.number_OG "12.8.73"
_space_group_magn.name_OG "C_2c2/m'"
_space_group_magn.point_group_number_Litvin "5.2.13"
_space_group_magn.point_group_name_UNI "2/m.1'"

loop_
_space_group_magn_transforms.id
_space_group_magn_transforms.Pp_abc
_space_group_magn_transforms.source
1 a,b,c;0,0,0 "BNS"
2 a,b,1/2c;0,1/2,-1/2 "OG"

loop_
_space_group_symop_magn_operation.id
_space_group_symop_magn_operation.xyz
1 x,y,z,+1 
2 -x,y,-z+1/2,+1 
3 -x,-y,-z,+1 
4 x,-y,z+1/2,+1 

loop_
_space_group_symop_magn_centering.id
_space_group_symop_magn_centering.xyz
1 x,y,z,+1 
2 x,y,z+1/2,-1 
3 x+1/2,y+1/2,z,+1 
4 x+1/2,y+1/2,z+1/2,-1 

loop_
_atom_site_label
_atom_site_type_symbol
_atom_site_symmetry_multiplicity
_atom_site_Wyckoff_symbol
_atom_site_fract_x
_atom_site_fract_y
_atom_site_fract_z
_atom_site_occupancy
_atom_site_fract_symmform
_atom_site_U_iso_or_equiv
_atom_site_adp_type
Mn1_1 Mn   4 c  0.00000  0.00000  0.00000  1.00000 0,0,0  0.00297(19)   Uiso
O1_1  O    4 b  0.00000  0.50000  0.75000  1.00000 0,0,0  0.00415(18)   Uiso

loop_
_atom_site_moment.label
_atom_site_moment.crystalaxis_x
_atom_site_moment.crystalaxis_y
_atom_site_moment.crystalaxis_z
_atom_site_moment.symmform
Mn1_1  4.610(10)  0.00000  0.00000 Mx,0,Mz 

#############################################
#
# NiO_C56_R_I-3.mcif
#
#############################################
# NiO 
# multi-k magnetic structure in MSG 148.20 R-3.1'_c[R-3], mL3+, OPD C56
_transition_temperature     525
_experiment_temperature     1.8

loop_
_irrep_id
_irrep_dimension
_irrep_small_dimension
_irrep_direction_type
_irrep_action
_irrep_modes_number
_irrep_presence
mL3+  6  2  special  primary   2 .
mL2+  4  1  .        secondary 2  yes

_parent_space_group.name_H-M_alt  "Fm-3m"
_parent_space_group.IT_number      225
_parent_space_group.transform_Pp_abc  "a,b,c;0,0,0"

loop_
_parent_propagation_vector.id
_parent_propagation_vector.kxkykz
k1 [1/2 1/2 1/2]
k2 [1/2,-1/2,1/2]
k3 [1/2,1/2,-1/2]

_parent_space_group.child_transform_Pp_abc  "-a-c,b+c,a+b-c;0,0,0"
_space_group_magn.transform_BNS_Pp_abc  "a,b,c;0,0,0"

_cell_length_a     5.90046
_cell_length_b     5.90046
_cell_length_c     14.42656
_cell_angle_alpha  90.00000
_cell_angle_beta   90.00000
_cell_angle_gamma  120.00000
_cell_volume       434.97586

_space_group_magn.number_BNS "148.20"
_space_group_magn.name_UNI "R-3.1'_c[R-3]"
_space_group_magn.name_BNS "R_I-3"
_space_group_magn.number_OG "148.4.1250"
_space_group_magn.name_OG "R_R-3"
_space_group_magn.point_group_number_Litvin "17.2.63"
_space_group_magn.point_group_name_UNI "-3.1'"

loop_
_space_group_magn_transforms.id
_space_group_magn_transforms.Pp_abc
_space_group_magn_transforms.source
1 a,b,c;0,0,0 "BNS"
2 -a,-b,1/2c;0,0,0 "OG"

loop_
_space_group_symop_magn_operation.id
_space_group_symop_magn_operation.xyz
1 x,y,z,+1 
2 -y,x-y,z,+1 
3 -x+y,-x,z,+1 
4 -x,-y,-z,+1 
5 y,-x+y,-z,+1 
6 x-y,x,-z,+1 

loop_
_space_group_symop_magn_centering.id
_space_group_symop_magn_centering.xyz
1 x,y,z,+1 
2 x,y,z+1/2,-1 
3 x+1/3,y+2/3,z+2/3,+1 
4 x+1/3,y+2/3,z+1/6,-1 
5 x+2/3,y+1/3,z+1/3,+1 
6 x+2/3,y+1/3,z+5/6,-1 

loop_
_atom_site_label
_atom_site_type_symbol
_atom_site_symmetry_multiplicity
_atom_site_Wyckoff_symbol
_atom_site_fract_x
_atom_site_fract_y
_atom_site_fract_z
_atom_site_occupancy
_atom_site_fract_symmform
_atom_site_U_iso_or_equiv
_atom_site_adp_type
Ni1_1 Ni  18 e  0.50000  0.00000  0.00000  1.00000 0,0,0 0.00041  Uiso
Ni1_2 Ni   6 a  0.00000  0.00000  0.00000  1.00000 0,0,0 0.00041  Uiso
O1_1  O   18 d  0.50000  0.00000  0.25000  1.00000 0,0,0 0.00145  Uiso
O1_2  O    6 b  0.00000  0.00000  0.25000  1.00000 0,0,0 0.00145  Uiso

loop_
_atom_site_moment.label
_atom_site_moment.crystalaxis_x
_atom_site_moment.crystalaxis_y
_atom_site_moment.crystalaxis_z
_atom_site_moment.symmform
Ni1_1 -0.655(3) -1.309(6)  1.619(7) Mx,My,Mz 
Ni1_2  0.00000  0.00000 -1.977(8) 0,0,Mz   

\end{verbatim}


\end{document}